\documentclass[5p]{elsarticle}
\usepackage{listings}
\usepackage{color}
  
\usepackage[cmex10]{amsmath}
\usepackage{graphicx}
\usepackage{epstopdf}
\usepackage{gensymb} 

\usepackage{amsthm}
\theoremstyle{definition}
\newtheorem{definition}{Definition}

\usepackage{algorithmicx}
\usepackage[ruled]{algorithm}
\usepackage{algpseudocode}
\vspace{-0.2cm}%
\alglanguage{pseudocode}

\usepackage{array}
\usepackage{booktabs}
\usepackage{multirow}

\DeclareMathOperator*{\argmin}{\arg\!\min}
\usepackage{threeparttable} 
\journal{Integration, The VLSI Journal}

\begin{document}

\begin{frontmatter}

\title{Detecting Recycled Commodity SoCs: \\Exploiting Aging-Induced SRAM PUF Unreliability}
%\title{Future Large-Scale Memristive Device Crossbar Arrays: Limits Imposed by Sneak-Path Currents on Read Operations}


\author[rvt]{Yansong Gao\corref{cor1}}
\ead{yansong.gao@adelaide.edu.au}

\author[els]{Hua Ma}
\ead{mary.ma@adelaide.edu.au}

\author[rvt]{Said F.~Al-Sarawi}
\ead{said.alsarawi@adelaide.edu.au}

\author[rvt]{Derek Abbott}
\ead{derek.abbott@adelaide.edu.au}

\author[els]{Damith C.~Ranasinghe}

%\ead[url]{http://www.elsevier.com}
\ead{damith.ranasinghe@adelaide.edu.au}

\cortext[cor1]{Corresponding author}
\address[rvt]{School of Electrical and Electronic Engineering,
The University of Adelaide, Adelaide, SA 5005, Australia}
%\address[focal]{School of Electrical and Computer Engineering, Royal Melbourne Institute of Technology, Victoria 3001, Australia}
\address[els]{School of Computer Science, The University of Adelaide, SA 5005, Australia}

%\author[mymainaddress]{Yansong Gao}
%\author[mysecondaryaddress]{Omid Kavehei}
%\author[mythirdaryaddress]{Damith C.Ranasinghe}
%\author[mymainaddress]{Said Al-Sarawi}
%\author[mymainaddress]{Derek Abbott}
%%%\ead[url]{www.elsevier.com}
%%%\author[mymainaddress,mysecondaryaddress]{Elsevier Inc}
%%%\ead[url]{www.elsevier.com}
%%
%%\author[mymainaddress]{School of Electrical and Electronic Engineering,\\
%%The University of Adelaide, Adelaide, SA, Australia 5000\corref{mycorrespondingauthor}}
%%\cortext[mycorrespondingauthor]{Corresponding author}
%%\ead{support@elsevier.com}
%%
%\address[mysecondaryaddress]{School of Electrical and Computer Engineering,\\Royal Melbourne Institute of Technology Victoria 3001, Australia}
%\ead{omid.kavehei@rmit.edu.au}
%\address[mythirdaryaddress]{School of Computer Science,\\
%The University of Adelaide, SA 5005, Australia}
%\ead{damith@cs.adelaide.edu.au}

\begin{abstract}
A physical unclonable function (PUF), analogous to a human fingerprint, has gained an enormous amount of attention from both academia and industry. SRAM PUF is among one of the popular silicon PUF constructions that exploits random initial power-up states from SRAM cells to extract hardware intrinsic secrets for identification and key generation applications. The advantage of SRAM PUFs is that they are widely embedded into commodity devices, thus such a PUF is obtained without a custom design and virtually free of implementation costs. A phenomenon known as `aging' alters the consistent reproducibility---reliability---of responses that can be extracted from a readout of a set of SRAM PUF cells.  
%Both applications favour consistently reproducible responses, however, natural response errors is non-trivial. Similar to create a PUF exploiting undesirable manufacturing randomness, SRAM PUF unreliability induced by aging can be exploited to detect recycled commodity devices requiring no additional cost to the device.
Similar to how a PUF exploits undesirable manufacturing randomness for generating a hardware intrinsic fingerprint, SRAM PUF unreliability induced by aging can be exploited to detect recycled commodity devices requiring no additional cost to the device. In this context, the SRAM PUF itself acts as an aging sensor by exploiting responses sensitive to aging. We use SRAMs available in pervasively deployed commercial off-the-shelf micro-controllers for experimental validations, which complements recent work demonstrated in FPGA platforms, and we present a simplified detection methodology along experimental results. We show that less than 1,000 SRAM responses are adequate to guarantee that both false acceptance rate and false rejection rate are no more than 0.001.
\end{abstract}


\end{frontmatter}
\section{Introduction}\label{Sec:Intro}

Electronic components are increasingly integrated and introduced into every domain of our lives. They are pervasively employed in Internet of Thing (IoT) devices such as wireless sensors in smart homes and health-care applications in civilian use cases to military and aerospace components in defense. However, over the past decade, counterfeit electronic components or integrated circuits (ICs) have flooded into every aspect of supply chains~\cite{ConterfeitReport}. Counterfeit ICs pose great concerns for: i) governments, threating national security or civilian safety due to their poor quality leading to lower performance or malfunctions that may result in critical system failures---e.g., transportation, hospital and power-station facilities, in addition, to tax revenue losses; ii) industry, they cause direct revenue loss and further ruin brand value; iii) consumers, they can induce potential safety concerns when they are employed in security or health critical applications due to the low quality and reliability issues~\cite{guin2014counterfeit}. 

Combating counterfeit ICs involves securing untrusted supply chains  resulting from the globalization of the semiconductor industry; one needs to trace, check and detect counterfeits along the supply chain within their life-cycles. Among various countermeasures, the physical unclonable function (PUF) is one promising lightweight hardware security primitive that assigns each IC with a unique identifier upon its creation, similar to fingerprints of humans~\cite{suh2007physical,jin2017secure,gao2015emerging,alvarez2016static}. Since PUF exploits manufacturing randomness, it is impossible for the counterfeiter to physically clone such instance-specific identifiers in the atom-by-atom level. Thus, the PUF is able to prevent counterfeiting ICs from several sources including cloned and overproduced ones. However, they were not considered to detect remarked and recycled counterfeit ICs~\cite{guin2014counterfeit} until recent work from~\cite{guo2016zero}. Extending PUF's functionality to detect remarked or recycled ICs is considerably valuable, as they contribute to more than 80\% of reported counterfeit incidents~\cite{guin2016design}.

Previous PUF applications focused on identification or authentication and key generation applications~\cite{gao2016obfuscated,yu2016pervasive}. In both, it is desirable for a PUF to regenerate the same response (output) when queried by the same challenge (input). However, in practice, the reliability of responses corresponding to certain challenges are affected by variations in environmental factors and aging effects. In typical PUF-based applications, for instance, cryptographic key generation requiring highly stable responses~\cite{maes2012pufky}, it is imperative to improve PUF reliability and correct potential bit errors prior to deriving a key. In PUF-based authentication applications \cite{ranasinghe2006confronting,suh2007physical}, it is still preferable to maximize reliability to reduce the number of response bits needed to uniquely identify a PUF instance from a large population and increase the complexity of modeling attacks by an adversary~\cite{lim2004extracting,ruhrmair2013puf,becker2015gap}.
 
In contrast, we take advantage of unavoidable unreliability of responses resulting from aging effects to provide a high degree of assurance to sense the period of aging experienced by PUF integrated ICs. In particular, we consider exploiting SRAM PUFs that are available in most commodity electronic systems or system on chips (SoCs), where neither additional area cost nor custom modification is required, to detect recycled commodity SoCs. The SRAM PUF is more suitable in this context in comparison with other popular silicon PUF structures such as Arbiter PUFs (APUF) and Ring Oscillator PUFs (ROPUFs)~\cite{herder2014physical,zhang2014survey,cao2015low} do requiring additional cost such as adding logic circuitry into existing electronic components using customized designs. Our work complement the recent work in~\cite{guo2016zero} utilizing SRAM PUFs to detect recycled devices demonstrated on FPGA platforms. We summarize our contributions below:
\begin{enumerate}
\item We evaluate and validate detection of recycled SoCs by using ubiquitously deployed micro-controllers that are commonly embedded with SRAM memories.
\item We develop a simplified aging sensitive response (ASR) selection methodology and detail how to systemically evaluate and quantify the detection capability. The detection is cost-free to the commodity SoCs since all the computations are left to the resource-rich verifier that carries out the detection. 
\item Our investigations with experimental results demonstrate that the aging-induced unreliability of SRAM PUFs in SoCs can effectively detect recycled SoCs with very high accuracy. Our ASR methodology allows to use less than 1,000 SRAM responses to ensure that both false rejection rate (FRR) and false acceptance rate (FAR) are less than $0.001$. In addition, experimental results validate that the detection accuracy increases with prolonged aging periods.
\end{enumerate}
\begin{figure}
 	\centering
 	\includegraphics[trim=0 0 0 0,clip,width=0.47\textwidth]{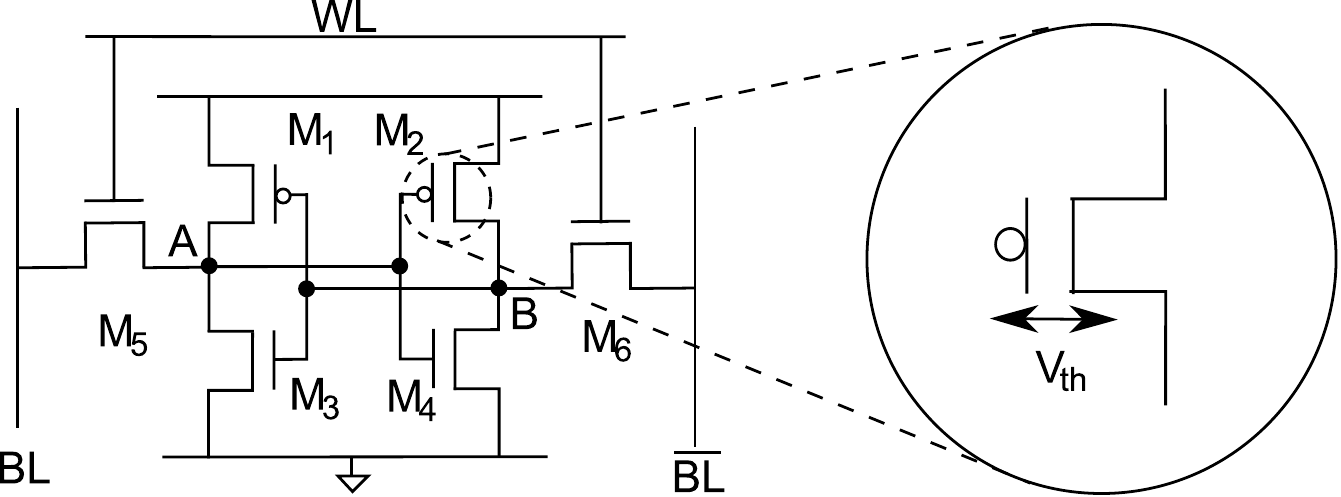}
 	\caption{SRAM cell \cite{holcomb2007initial}. $ V_{\rm th} $ difference in the transistors results into repeatable random power-up states either in `1' or `0'.}
 	\label{fig:ROPUF}
 \end{figure}
The rest of the paper is organized as follows. Related work is introduced in Section~\ref{Sec:Relate}. In Section \ref{Sec:DetecMethod}, we detail the simplified ASR selection methodology and how to systematically evaluate the detection capability. The simplified response selection approach during the provisioning phase is introduced in order to improve the detection efficiency in the recycled hardware detection phase by employing responses that are more sensitive to aging effects. Then comprehensive experimental results from off-the-shelf commodity microcontrollers embedded SRAM PUFs are given in Section~\ref{Sec:Experiments}. In Section~\ref{Sec:Conclusion} we conclude this article.

\section{Background and Related Work}\label{Sec:Relate}
\subsection{SRAM PUF}\label{Sec:SRAMPUF}
Unlike the other two popular silicon PUF constructions, ROPUFs and APUFs that exploit time delay differences~\cite{suh2007physical} to extract secrets, SRAM PUFs~\cite{su20071,holcomb2007initial} leverage the threshold voltage $ V_{\rm th} $ mismatch between two cross-coupled inverters of a SRAM cell resulting from manufacturing randomness. As a memory cell, a write operation forces the SRAM cell to transition into one of two digital states, e.g., `0' or `1'. When a cell is powered up or no write operation is occurred, the SRAM cell tends to prefer a repeatable power-up state---also referred as a response---either being `1' (AB=01) or `0' (AB=10). As an example, if the $ V_{\rm th, P_1} $ is slightly smaller than $ V_{\rm th, P_2} $, at power-up, M$_1 $ starts conducting before M$ _2 $, thus, $ A=1 $. This in turn prevents M$_2 $ switching on. As a consequence, the SRAM cell at power-up prefers to be `0' (AB=10). Larger $ |V_{\rm th, P_1} - V_{\rm th, P_2}| $ leads to more repeatable power-up states or more reliable responses when the cell is used to regenerate the response. Such a repeatable power-up state differs from cell to cell and chip to chip as well, thus, a map of the power-up states of a set of SRAM cells can be treated as a unique identifier, or the SRAM memory array can be treated as a PUF. In particular, the readout SRAM power-up state is referred to as the {\it response}, while the address of the SRAM cell is referred to as the {\it challenge}.

\subsection{Sensing Aging with SRAM PUFs}
However, some of SRAM PUF responses are not reproducible due to that fact that the $ V_{\rm th} $ difference of a selected cell is not dominant in the presence of noise from environmental factors such as supply voltage and temperature variations and aging effects. In elementary PUF-oriented identification and authentication applications, those unreliable responses are undesirable. In contrast, and just as undesirable fabrication randomness is extracted to create instance-specific PUFs to derive a physical inseparable trust anchor for a hardware device, a PUF response's sensitivity to environmental factors and aging can also be utilized to secure sensing. In this context, sensing functionality is derived from a PUF and the PUF lends itself as a sensor to guarantee the veracity of sensed data~\cite{rosenfeld2010sensor,ruhrmair2015virtual,baby2016finite,guo2016zero,gao2017pufsensor}. 

%However, some of SRAM PUF responses are not repeatable due to the $ V_{\rm th} $ difference of a selected cell not being dominant in the presence of noise from environmental factors such as supply voltage and temperature variations. In elementary PUF-oriented identification and authentication applications, those unreliable responses are undesirable. In contrast, and just as undesirable fabrication randomness is extracted to create instance-specific PUFs to dervie a physicaly inseparable trust anchor into a hardware device, a PUF response's sensitivity to environmental factors can also be utilized to secure sensing. In this context, sensing functionality is dervied from a PUF and the PUF itself acting as a sensor to guarantee the veracity of sensed data~\cite{rosenfeld2010sensor,ruhrmair2015virtual,baby2016finite,guo2016zero,gao2017pufsensor}. 

Detecting recycled devices using SRAM PUFs by considering those PUF responses sensitive to aging effects was recently received attention and initially investigated in~\cite{guo2016zero}. Guo {\it et al.} used SRAM cells in FPGA platforms for experimental validations. We complement this initial investigation using SRAM memory in pervasively deployed off-the-shelf micro-controllers as they are commonly deployed in many SoCs ranging from home appliances to various sensors in the Internet of Things (IoT) era. We present a simpler methodology of selecting aging sensitive response bits and detail a systematic approach to evaluate and quantify recycled SoC detection capability supported by experimental data.

\subsection{SRAM Aging}
Silicon ICs performance deteriorates gradually over time attributing to various factors such as hot carrier injection (HCI), time-dependent dielectric breakdown (TDDB) and bias temperature instability (BTI)~\cite{maes2014countering,guo2016zero}. The negative BTI (NBTI) plays dominant aging effect in modern ICs, especially for switched-on pMOS transistors~\cite{maes2014countering}.

The NBTI effect increases the threshold voltage of pMOS transistors when a transistor is `on'. Considering the example in Section~\ref{Sec:SRAMPUF} where the SRAM is powered up without a write operation. Consider that we already knew AB=10, where the M$_1 $ is `on' and experiences a gradually increased $ V_{\rm th, P_1} $ due to the NBTI, while the $ V_{\rm th, P_2} $ remains or changes negligibly with respect to $ V_{\rm th, P_1} $. Hence, over time, $ V_{\rm th, P_1} > V_{\rm th, P_2} $. As a consequence, the regenerated responses over the life of such cells tend to shift from being reliably generated `0' to `1'. Though anti-aging strategies are possible~\cite{maes2014countering}, its expensive time and monetary cost prohibit a counterfeiter to do so, especially for low-end ICs. The bit flipping over time caused by the response sensitivity to aging is undesirable for conventional PUF applications, but can be exploited to detect recycled commodity SoCs widely embedded with SRAM memories.

\section{Detection Methodology}\label{Sec:DetecMethod}
Only a small fraction of SRAM responses are sensitive to aging over time; we will experimentally show this in Section~\ref{Sec:Experiments}. In other words, most response bits are actually reproduced consistently across a wide range of operating conditions and aging effects. Such response bits are desirable for elementary PUF authentication and key generation applications, but cannot be utilized for sensing aging as they are invariant to aging effects. 

Therefore, we need to first efficiently select and determine those ASRs during the provisioning phase---after the SoCs are fabricated but prior being delivered through a(n) (insecure) supply chain---to facilitate detection of recycled SoCs in the detection phase later on. 
%In essence, this requires regenerating response bits to determine those exhibiting higher errors rates after aging effects in comparison to error rates resulting from environmental variations before aging where a responses reliability before aging is used as a reference. However, such an approach is very inefficient and cumbersome.
Hence, we develop a simplified methodology of selecting and determining ASRs followed by elaborating on how to systematically evaluate the detection capability utilizing those ASRs.

%One approach would be to employ randomly selected challenges and evaluate responses for their sensitivity to ageing. 

%that solely resulting from noise when the response evaluated before aging is used as the reference. 

Before delving into detailed descriptions, we give a number of useful definitions to ease the following descriptions, especially the systematic detection capability evaluations.

\subsection{Preliminaries}
\begin{definition}{\bf InterA-distance.}\label{Def:InterAge-distance}
The interA-distance is a random variable describing the distance between two PUF responses $ {\bf R}^{\rm PreA}, {\bf R}^{\rm PostA} $ produced before aging and after aging by applying the same challenge---address in case of a SRAM PUF---to the same PUF, hence,

\begin{equation}\label{Eq:InterAge}
 D_{\rm interA}={\rm dist}({\bf R}^{\rm PreA}, {\bf R}^{\rm PostA})
 \end{equation}
 where $ {\bf R}^{\rm PreA}, {\bf R}^{\rm PostA} $ are two responses generated before and after aging by applying the same challenge to the same PUF.
\end{definition}

\begin{definition}{\bf IntraA-distance.}\label{Def:IntraAge-distance}
The intraA-distance is a random variable describing the distance between two PUF responses $ {\bf R}^{\rm A}, {\bf R}^{\rm A^{\prime}} $ re-evaluated on the same PUF, using the same challenge before aging.

\begin{equation}\label{Eq:IntraPQ}
 D_{\rm intraA}={\rm dist}({\bf R}^{\rm A}, {\bf R}^{\rm A^{\prime}})
\end{equation}
 where $ {\bf R}^{\rm A}, {\bf R}^{\rm A^{\prime}} $ are two responses obtained from the same PUF using the same chosen challenge before aging.
\end{definition}
 
The dist(.;.) can be any well-defined and appropriate distance metric over the responses. In this paper, responses are always bit vectors and the used distance metric is Hamming distance (HD) or fractional Hamming distance formally defined below:

\begin{definition}{\bf Hamming distance.}
 	For bit vectors $ {\bf X}_1 $ and $ {\bf X}_2 $ with the same length $ l $, the HD between them is defined as:
 	\begin{equation}\label{eq:HD}
 	{f_{\rm HD}}({\bf X}_1, {\bf X}_2)=\sum_{i=1}^{l} {\bf X}_1\oplus {\bf X}_2.
 	\end{equation}
 \end{definition}
 
 \begin{definition}{\bf Fractional Hamming distance.}
 	Built upon Eq.~(\ref{eq:HD}), the fractional Hamming distance (FHD) is defined as:
 	\begin{equation}\label{eq:FHD}
 	f_{\rm FHD}({\bf X}_1, {\bf X}_2)=\frac{f_{\rm HD}({\bf X}_1, {\bf X}_2)}{l}.
 	\end{equation}
 \end{definition}
 
Readers who are familiar with PUFs will notice that the definition of the interA-distance is similar to the inter-distance of PUFs that measures the difference between two responses from two distinct PUF instances given the same challenge. The difference is that the interA-distance is evaluated across differing aging periods subject to the same PUF instance, the inter-distance is, however, evaluated across different PUF instances.
 
The intraA-distance is similar to the intra-distance of PUF responses that measures the difference between two responses reproduced from two distinct evaluations by applying the same challenge to the same randomly chosen PUF instance. The main difference is that the intra-distance does not consider the source of aging, it simply treats any environmental fluctuation, e.g., supply voltage, temperature and also aging effects as noise sources. However, in this work, we are able to finely fix the supply voltage and temperature, only thermal noise is treated as a noise source. The aging effects is not a noise source but is exploited to detect aging devices.
 
Similar to the inter-distance and intra-distance distribution of PUFs explained in detail in~\cite{roel2012physically}, both of the interA-distance and intraA-distance can be assumed to follow a binomial distribution $ B(n,p) $. The binomial probability estimator of interA-distance and intraA-distance distributions are referred to as $ \hat{p}_{\rm interA} $ and $ \hat{p}_{\rm intraA} $, respectively. In general, the $ \hat{p}_{\rm interA} $ is the probability that $ {\bf R}^{\rm PreA}\ne {\bf R}^{\rm PostA} $, see Definition~\ref{Def:InterAge-distance}, and the $ \hat{p}_{\rm intraA} $ is the probability that $ {\bf R}^{\rm A}\ne {\bf R}^{\rm A^{\prime}} $, see Definition~\ref{Def:IntraAge-distance}. 

\subsection{Detecting Capability}\label{Sec:DetectCap}
  \begin{figure}[h]
  	\centering
  	\includegraphics[trim=0 0 0 0,clip,width=0.47\textwidth]{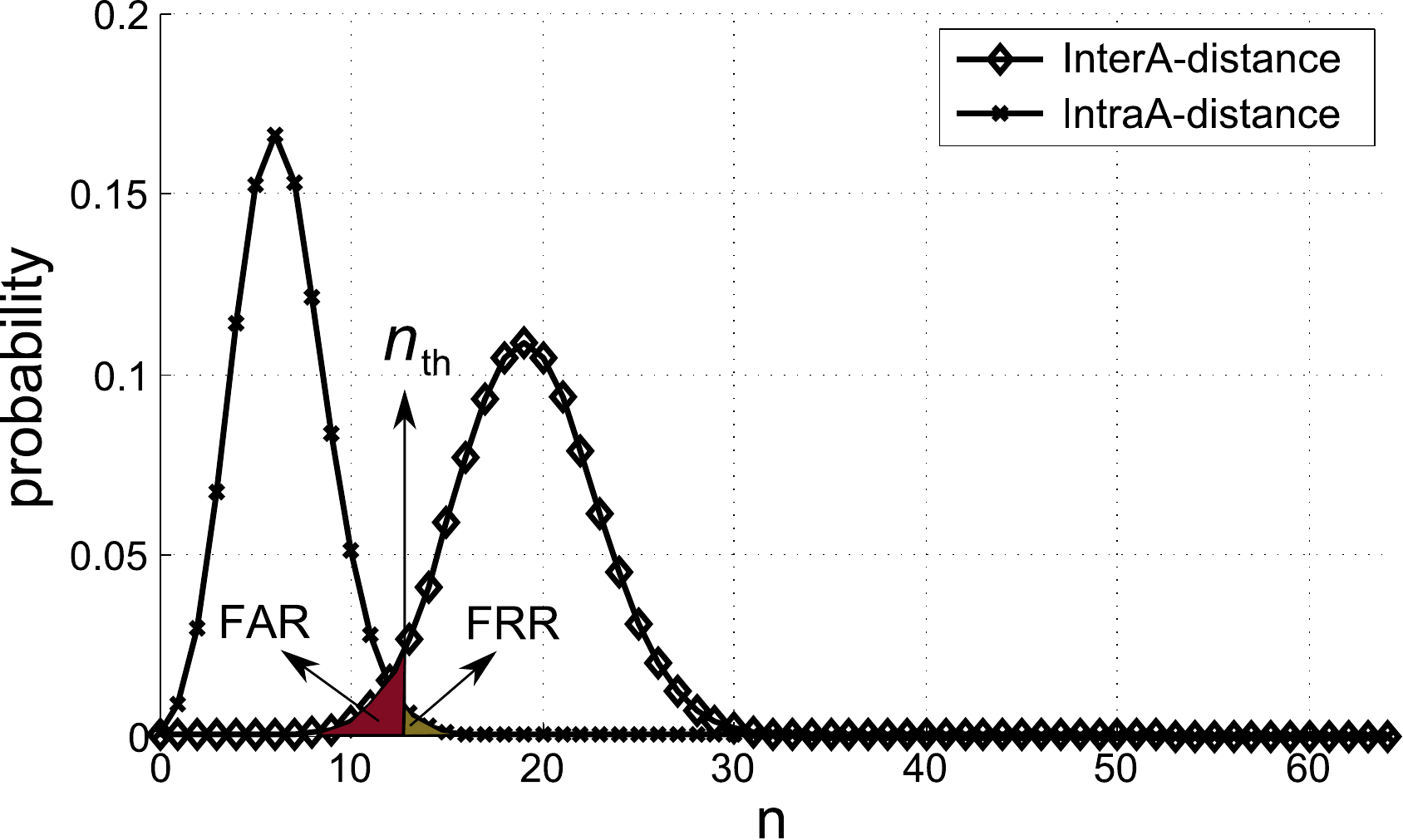}
  	\caption{Illustration of distribution of interA-distance and intraA-distance for a 64-bit response.}
  	\label{FARFRR}
  \end{figure}
Clearly one single challenge-response pair or CRP is not able to correctly detect an aged device. We need to use multiple response bits or a number of CRPs to minimize the error for: i) mistakenly accepting a response from a PUF that has not undergone aging, referred as false acceptance rate (FAR); and ii) falsely rejecting an  authentic response when it a regenerated from an aged PUF, referred as false rejection rate (FRR). It is imperative to minimize both FAR and FRR in practice. More generally, FAR stands for the probability of incorrectly regarding a new device as an aged one. While FRR stands for the probability of an aged device being falsely rejected as a new device. 
 
These two undesirable errors are illustrated in Fig.~\ref{FARFRR}. The right tail of the intraA-distance distribution indicates the FRR, while the left tail of the interA-distance distribution depicts the FAR. When the length of response bits or the number of CRPs, $ n $, and the threshold $ n_{\rm th} $ used for achieving a desirable FAR and FRR, and considering that both interA-distance and intraA-distance follow a binomial distribution, FAR and FRR can be formally expressed following work in~\cite{lim2005extracting,roel2012physically}:

\begin{equation}\label{eq:FRR}
 {\rm FRR} = 1 - \sum\limits_{i=0}^{{n}_{\rm th}}{\binom{n}{i}}({ {{\hat p}_{\rm intraA}}})^i{(1-{{ {\hat p}_{\rm intraA}}})^{(n-i)}},
 \end{equation}
 
 \begin{equation}\label{eq:FAR}
 {\rm FAR} =\sum\limits_{i=0}^{{n}_{\rm th}}{\binom{n}{i}}({ {\hat p}_{\rm interA}})^i{(1-
 	{{\hat p}_{\rm interA}})^{(n-i)}}.
 \end{equation}
 
Based on (\ref{eq:FRR}) and (\ref{eq:FAR}), we can see that the FRR and FAR depend on the $ \hat{p}_{\rm intraA} $ and $ \hat{p}_{\rm interA} $, the threshold $ {\rm n}_{\rm th} $, and the number of employed CRPs $ n $. For example, supposing $ n $ is 64 as shown in Fig.~\ref{FARFRR}, a large $ n_{\rm th} $ benefits the false rejection rate but aggravates the false acceptance rate, and vice versa for a small $ n_{\rm th} $. We want to minimize both FAR and FRR in practice. There exists a threshold value to make both FAR and FRR equal. We refer this interested threshold value as {\it equal error threshold}, termed $ n_{\rm EER} $. Consequentially, when both error rates are equal, we refer this equal rate as {\it equal error rate} (EER) following Roel's work~\cite{roel2012physically}. For a discrete distribution, there may not be an $ n_{\rm EER} $ for which FAR is equal to FRR, and in that case, $ {n}_{\rm EER} $ and EER are defined as in~\cite{roel2012physically}:
 \begin{equation}
 {n}_{\rm EER} = \argmin_{{n}_{\rm th}} \{ {\rm max}\{{\rm FAR}({n}_{\rm th}),{\rm FRR}({n}_{\rm th}) \}\},
 \end{equation}
 \begin{equation}
 {\rm EER} =  \max \{{\rm FAR}({n}_{\rm EER}),{\rm FRR}({n}_{\rm EER})\}.
 \end{equation}
 Given binomial probability estimator ${\hat p}_{\rm interA} $ and $ {\hat p}_{\rm intraA} $, the task is to find minimal number of CRPs, $ n $, for ensuring an acceptable $ \rm EER $ that meet desired requirements. 
% We will quantize the $ n $ in Section~\ref{Sec:Experiments} based on experimentally evaluated ${\hat p}_{\rm interA} $ and $ {\hat p}_{\rm intraA} $. 

%minimizing both FAR and FRR
%---will be experimentally validated in Section~\ref{Sec:Experiments},

To increase the capability of distinguishing  recycled devices from new ones and minimize both FAR and FRR, it is imperative to increase the difference between  $ \hat{p}_{\rm intraA} $ and $ \hat{p}_{\rm interA} $. We can visually observe this in Fig.~\ref{FARFRR}. For example, when the interA-distance distribution shifts to right and intraA-distance distribution keeps same, it is clear that both FAR and FRR will be reduced as the overlapped area becomes small. Therefore, we introduce an approach to select SRAM responses that are of higher sensitivity to aging to increase the difference between $ \hat{p}_{\rm intraA} $ and $ \hat{p}_{\rm interA} $.

\subsection{Selecting ASRs}
It has been shown when a SRAM cell is under high temperature,  $ V_{\rm th} $ increases in a similar manner to that caused by aging~\cite{guo2016zero}. Therefore, during provisioning phase, the SRAM PUF responses can be re-evaluated under room temperature (RT) and high temperature (HT), respectively, to select aging sensitive responses (ASRs). Notably, the high temperature setting is only necessary during the provisioning phase and is not required during the detection phase. The ASR selection follows {\bf Algorithm~\ref{Algorithm:SelASR}}.

\begin{algorithm}[h]
	\small
	\caption{Selecting ASRs}
	\label{Algorithm:SelASR}
	\begin{algorithmic}[1]
		\Procedure{$\mathbf{selection}$~} {PUF, RT, HT}
		\For{$ i=1:N $}
		\State generating response r$ _{\rm RT_i} $ under RT using PUF;
		\EndFor
		\For{$ i=1:N $}
		\State generating response r$ _{\rm HT_i} $ under HT using PUF;
		\EndFor
		%\State 
		\If {(all r$ _{\rm RT_i} $ same) \&\& (all r$ _{\rm HT_i} $ same)
		 \&\& (r$ _{\rm RT} $$ \ne $r$ _{\rm HT} $)}
		\State select aging sensitive response ${\rm r}$;
		\State \textbf{return}
		\Else
		\State discard response ${\rm r}$;
		\State \textbf{return}
		\EndIf
		\EndProcedure
		\Statex
	\end{algorithmic}
	\vspace{-0.4cm}%
\end{algorithm}

The proposed ASR selection method is straightforward and simpler in comparison with~\cite{guo2016zero}. During the provisioning phase, the response r is regenerated $ N $ times under RT and HT respectively. The r is selected as an ASR when all regenerated r are same under RT and HT, respectively, but exhibit {\it opposite} values. For example, the regenerated r exhibits `1' for all $ N $ evaluations under RT and `0' for all $ N $ evaluations under HT. Then this r is selected as a ASR. Otherwise, it is discarded and will not be utilized for detecting aging SoCs in the afterward detection phase.

When ASRs are selected, there $ \hat{p}_{\rm intraA} $ and $ \hat{p}_{\rm interA} $ can be heuristically evaluated. We assume $ \hat{p}_{\rm intraA} $ is less than $ \hat{p}_{\rm interA} $, and this is true as we will show in Section~\ref{Sec:Experiments}.
%The heuristically evaluated $ \hat{p}_{\rm intraA} $ and $ \hat{p}_{\rm interA} $ based on a very small number of SRAM PUFs can be used to derive
%Please noting that $ \hat{p}_{\rm interA} $ is evaluated under the same nominal supply voltage and room temperature. This is easy to be set during the detection phase, where noting HT setting is unnecessary.

\section{Experimental Results}\label{Sec:Experiments}
\subsection{Experiment Setup}
\begin{figure}[h]
	\centering
	\includegraphics[trim=0 0 0 0,clip,width=0.47\textwidth]{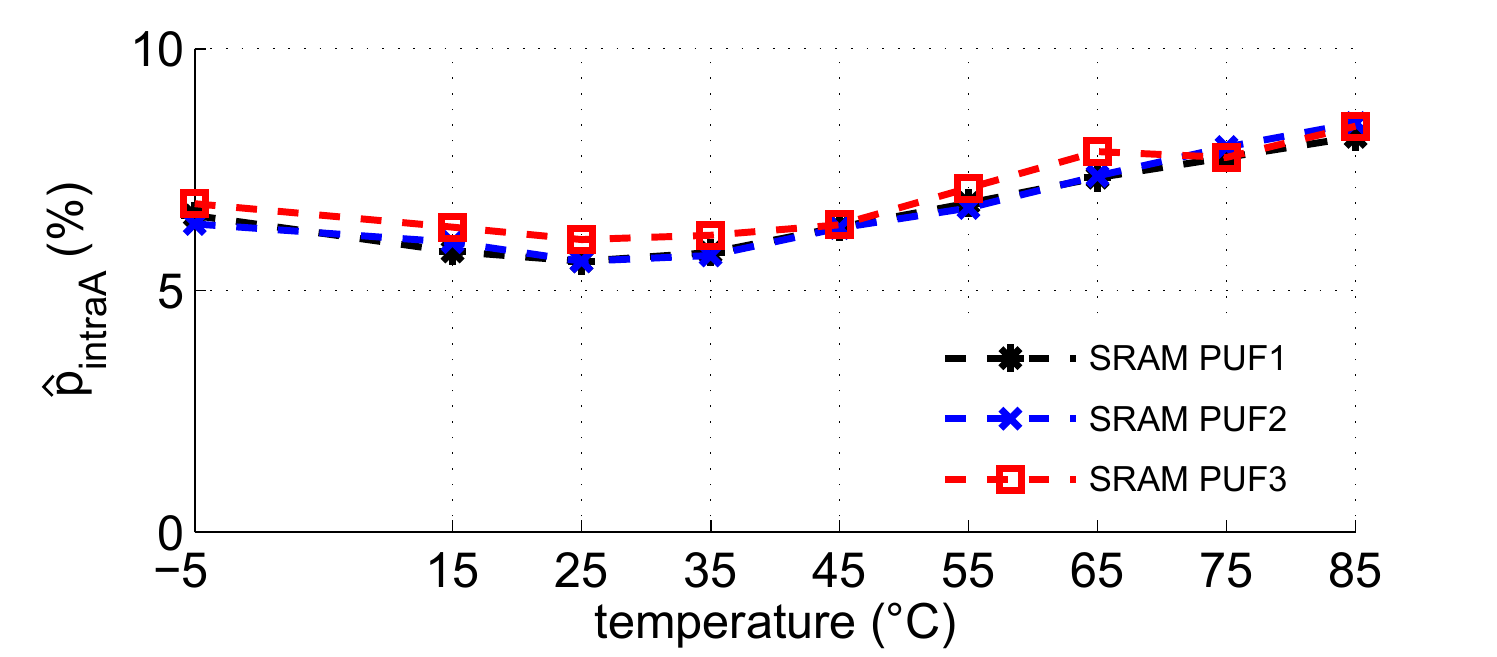}
	\caption{$ {\hat p}_{\rm intraA} $ of three SRAM PUFs across three microcontrollers under nine temperature corners, reference temperature is $ 25\celsius $.}
	\label{fig:SRAMPUFBER}
\end{figure}

SRAM PUF CRP dataset is collected from three chipKIT Pro MX7 microcontroller boards. From each board, we read power-up states from 262,144 SRAM cells as SRAM PUF responses. The nominal power supply voltage is 3.25~V. We are able to change the voltage from 3.125~V to 3.50~V. We found that the voltage, however, has negligible effects on the SRAM PUF reliability under test, which agrees with other experimental results~\cite{roel2012physically}. Therefore, we focus on SRAM PUF reliability performance as shown in Fig.~\ref{fig:SRAMPUFBER} that is $ {\hat p}_{\rm intraA} $ before aging under nine different temperature corners: $ -5\celsius $, $ 15\celsius $, $ 25\celsius $, $ 35\celsius $, $ 45\celsius $, $ 55\celsius $, $ 65\celsius $, $ 75\celsius $, $ 85\celsius $. The room temperature $ 25\celsius $ is treated as the nominal or reference corner. We are mostly interested in the $ {\hat p}_{\rm intraA} $ under RT, which is approximately 6\% as shown in Fig.~\ref{fig:SRAMPUFBER}. 

To test the aging influence on the SRAM PUF response's reliability, we put the microcontroller board in the oven of $ 80\celsius $ to accelerate the aging. For expected NBTI aging, the {\it acceleration factor} (AF) is expressed~\cite{maes2014countering}:
\begin{equation}
AF=(\frac{V_{\rm stress}}{V_{\rm nominal}})^{\frac{\alpha}{\rm m}}\cdot {\rm exp}\big(\frac{E_{\rm aa}}{k}\cdot(\frac{1}{T_{\rm stress}}-\frac{1}{T_{\rm nominal}})\cdot{1\over m}\big),
\end{equation}
where the parameters setting are: the gate voltage exponent $ \alpha=3.5 $; the time exponent $ m=0.25 $; the apparent activation energy $ E_{\rm aa} =-0.02eV$; and Boltzmann's constant $ k=8.62\times 10^{-5}eV/K $. We only consider temperature resulted stress, where $ V_{\rm stress}=V_{\rm nominal} $, $ T_{\rm stress}=80\celsius, T_{\rm nominal}=25\celsius $. As a consequence, we are able to obtain AF=11.03. 
\begin{figure}[h]
	\centering
	\includegraphics[trim=0 0 0 0,clip,width=0.47\textwidth]{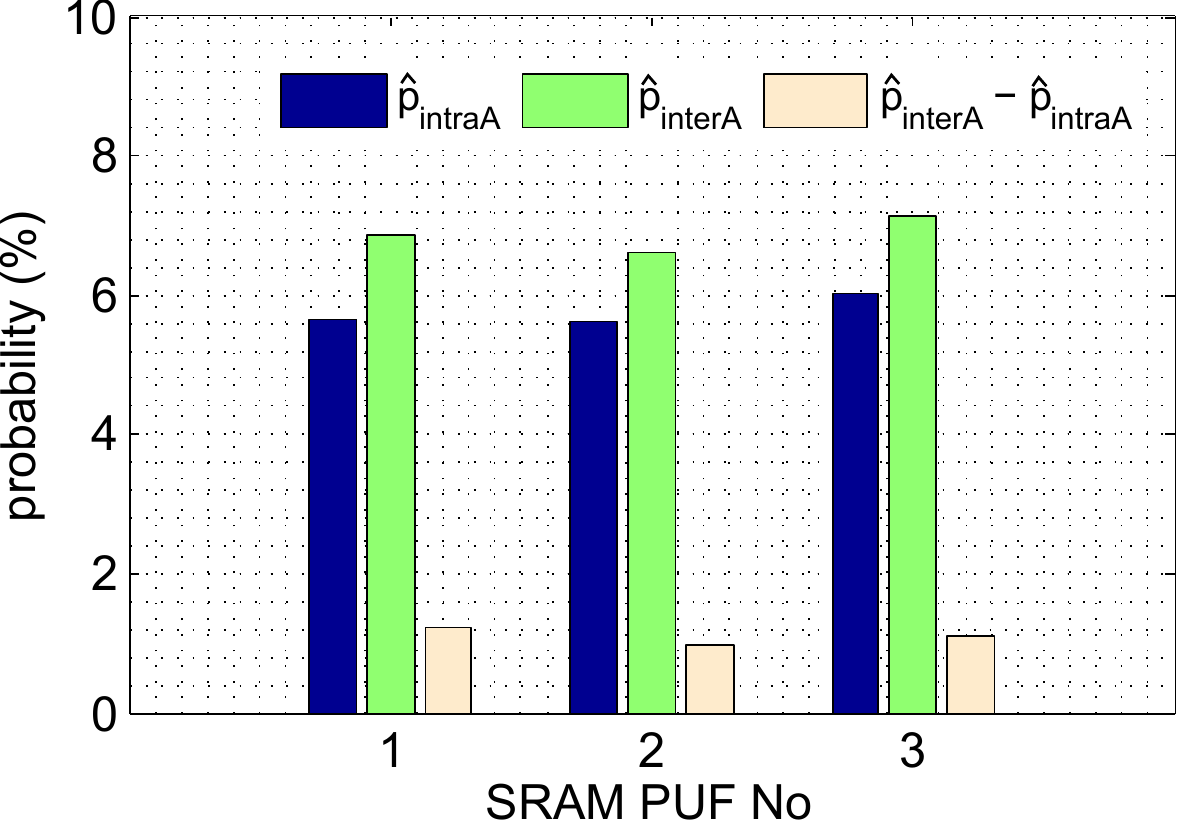}
	\caption{$ {\hat p}_{\rm intraA} $ and $ {\hat p}_{\rm interA} $ of three SRAM PUFs evaluated under nominal supply voltage, 3.25~V, and nominal or room temperature, $ 25\celsius $.}
	\label{fig:PInterIntraNASR}
\end{figure}

After 48 hours of accelerated aging that is equal to 22.1 days of effective NBTI device aging under normal working conditions in the field, we calculate the $ {\hat p}_{\rm interA} $ under RT using a strategy of randomly selecting responses. Results in Fig.~\ref{fig:PInterIntraNASR} imply that $ {\hat p}_{\rm interA} $ is only slightly higher than $ {\hat p}_{\rm intraA} $. More specifically, the difference between $ {\hat p}_{\rm interA}$ and ${\hat p}_{\rm intraA} $ is only around 1\%. This indicates that only a small fraction of responses are sensitive to aging. We can see from our analyses in Section~\ref{Sec:DetectCap} that using a random response selection strategy for recycled SoCs detection is cumbersome. 

Next, we first implement the ASR selection approach outlined in {\bf Algorithm~\ref{Algorithm:SelASR}} and then demonstrate the significantly improved difference between $ {\hat p}_{\rm interA}$ and ${\hat p}_{\rm intraA} $ that consequently facilitates the detection capability.
\subsection{ASR Detection Capability Results}
\begin{figure}
	\centering
	\includegraphics[trim=0 0 0 0,clip,width=0.47\textwidth]{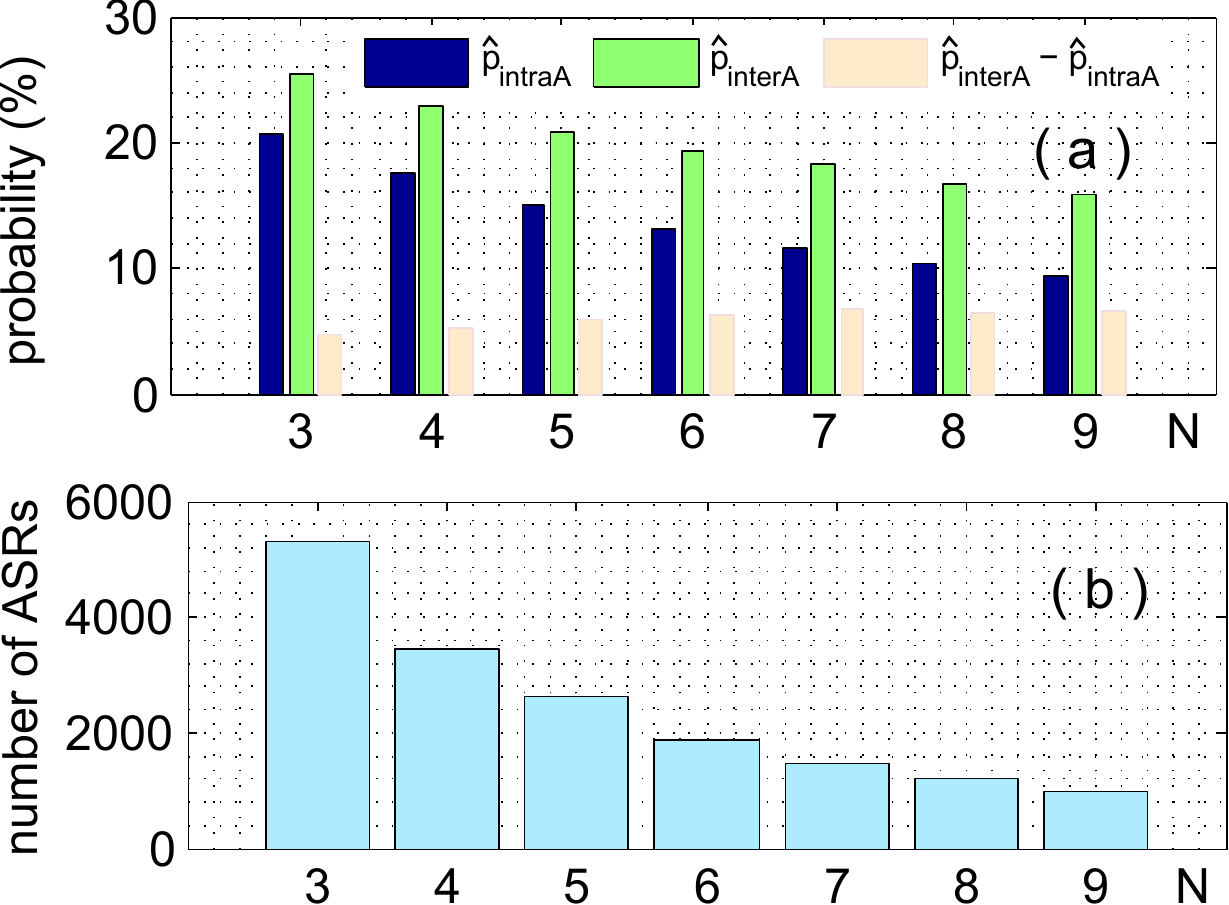}
	\caption{(a) Mean of $ {\hat p}_{\rm intraA} $, $ {\hat p}_{\rm interA} $ and $ {\hat p}_{\rm interA} -{\hat p}_{\rm interA} $ as a function of $ N $. A larger $ {\hat p}_{\rm interA} -{\hat p}_{\rm interA} $ and, at the same time, a smaller $ {\hat p}_{\rm intraA} $ are desirable in practice, which is achieved by increasing $ N $. (b) Average of number of selected ASRs as a function of $ N $.}
	\label{fig:PInterIntraYASR}
\end{figure}
We apply the ASR selection according to {\bf Algorithm~\ref{Algorithm:SelASR}}. Noting that both $ {\hat p}_{\rm intraA} $ and $ {\hat p}_{\rm interA} $ are a function of $ N $, which is number of a response reevaluated under a given RT and HT. The purpose of the selection process is to increase the difference between $ {\hat p}_{\rm interA} $ and $ {\hat p}_{\rm intraA} $ whilst also making sure that the $ {\hat p}_{\rm intraA} $ is small as well. Mean of $ {\hat p}_{\rm interA} $, $ {\hat p}_{\rm intraA} $, and $ {\hat p}_{\rm interA}- {\hat p}_{\rm intraA} $ as a function of $ N $ settings for those selected ASRs are depicted in Fig.~\ref{fig:PInterIntraYASR} (a). We can see that $ {\hat p}_{\rm intraA} $, when ASR is implemented, is always larger than the $ {\hat p}_{\rm intraA} $ of around 6\% without ASR selection, see Fig.~\ref{fig:PInterIntraNASR}, this is because the ASRs are also tending to be erroneous when they are regenerated under RT before aging. However, the $ {\hat p}_{\rm interA} $ is increased faster, therefore, larger $ {\hat p}_{\rm interA}- {\hat p}_{\rm intraA} $ is achieved. In addition, the $ {\hat p}_{\rm intraA} $ decreases as the $ N $ increases with slightly improved $ {\hat p}_{\rm interA}- {\hat p}_{\rm intraA} $. Overall, as we shall see in Table~\ref{tab:AuthenTab}, a larger $ N $ yields a higher detection capability.

In Fig.~\ref{fig:PInterIntraYASR} (b), the number of selected ASRs out of 262,144 responses are depicted. We can expect that the number of ASRs to decrease as $ N $ increases because less number of responses are able to satisfy the selection criterion in {\bf Algorithm~\ref{Algorithm:SelASR}}. Therefore, a larger $ N $ leads to less number of selected ASRs but higher sensitivity to aging for those selected ASRs. 
\begin{table*} 
 \centering
 \caption{Quantitative evaluation of necessary bit length of the response for successful detection under different ${\hat p}_{\rm interA} $ and $ {\hat p}_{\rm intraA} $ that are determined by $N $.}
 	\begin{tabular}{c c c c|| p{0.5cm} p{0.5cm} p{0.5cm} p{0.8cm} || p{0.5cm} p{0.5cm} p{0.5cm} p{0.8cm} || p{0.5cm} p{0.5cm} p{0.5cm} p{0.8cm}} % The final bracket specifies the number of columns in the table along with left and right borders which are specified using vertical bars (|); each column can be left, right or center-justified using l, r or c. To specify a precise width, use p{width}, e.g. p{5cm}
 		\toprule % Top horizontal line
 		\toprule % Top horizontal line
 		& & &\multicolumn{4}{r}{EER $ < 10^{-2} $} & \multicolumn{4}{r}{EER $ < 10^{-3} $}& \multicolumn{4}{r}{EER $ < 10^{-4} $} \\% Amalgamating several columns into one cell is done using the \multicolumn command as seen on this line
 		\cmidrule(l){5-8} \cmidrule(l){9-12} \cmidrule(l){13-16}% Horizontal line spanning less than the full width of the table - you can add (r) or (l) just before the opening curly bracket to shorten the rule on the left or right side
 		$ N $ & $ {\hat p}_{\rm intraA} $ & ${\hat p}_{\rm interA} $&$ {\hat p}_{\rm interA}-{\hat p}_{\rm intraA} $ & $ n $ & $ {\rm n}_{\rm EER} $ & FAR$ ^* $ & FRR$ ^* $ &$ n$ & $ {\rm n}_{\rm EER} $ & FAR$ ^* $ & FRR$ ^* $ & $ n $ & $ {\rm n}_{\rm EER} $ & FAR$ ^* $ & FRR$ ^* $  \\ % Column names row
 		\midrule % In-table horizontal line
 		3 & 20.70\% & 25.45\%& 4.75 \% & 1706 & 393 & $ -2.01 $ & $ -2.01 $ & 3005 & 692 & $ -3.01 $ & $ -3.00 $ & 4347 & 1001 & $ -4.01 $ & $ -4.01 $ \\ % Content row 1
 		4 & 17.55\% & 22.84\%& 5.29 \% & 1251 & 252 & $ -2.01 $ & $ -2.00 $ & 2191 & 441 & $ -3.00 $ & $ -3.01 $ & 3171 & 638 & $ -4.00 $ & $ -4.01 $ \\ % Content row 2
 		5 & 14.98\% & 20.87\%& 5.89 \% & 914 & 163 & $ -2.01 $ & $ -2.00 $ & 1611 & 287 & $ -3.01 $ & $ -3.00 $ & 2330 & 415 & $ -4.00 $ & $ -4.01 $ \\ % Content row 3
 		6 & 13.07\% & 19.32\%& 6.25 \% & 746 & 120 & $ -2.00 $ & $ -2.01 $ & 1314 & 211 & $ -3.01 $ & $ -3.00 $ & 1906 & 306 & $ -4.00 $ & $ -4.01 $ \\ % Content row 4
 		7 & 11.54\% & 18.28\%& 6.74 \% & 603 & 89 & $ -2.02 $ & $ -2.02 $ & 1052 & 155 & $ -3.00 $ & $ -3.01 $ & 1528 & 225 & $ -4.01 $ & $ -4.02 $ \\ % Content row 5
 		8 & 10.30\% & 16.73\%& 6.43 \% & 606 & 81 & $ -2.01 $ & $ -2.01 $ & 1065 & 142 & $ -3.01 $ & $ -3.00 $ & 1546 & 206 & $ -4.00 $ & $ -4.01 $ \\ % Content row 6
 		9 & 9.26\% & 15.78\%& 6.52 \% & 551 & 68 & $ -2.01 $ & $ -2.00 $ & 974 & 120 & $ -3.01 $ & $ -3.04 $ & 1406 & 173 & $ -4.01 $ & $ -4.03 $ \\ % Content row 7 		 		 		
 		\bottomrule % Bottom horizontal line
 		%\bottomrule % Bottom horizontal line
 	\end{tabular}
 	\begin{tablenotes}  
 		\item[a] Note: the $ ^* $ symbol indicates $ \log_{10}({\cdot}) $ of the value.
 		
 	\end{tablenotes}  
 	\label{tab:AuthenTab} % A label for referencing this table elsewhere, references are used in text as \ref{label}
 \end{table*}

In Table.~\ref{tab:AuthenTab}, we give results of quantitatively evaluations of $ n $---minimal bit length of the response to meet the EER, and $ n_{\rm th} $ or $ n_{\rm EER} $ of SRAM PUF being used to detect recycled SoCs under different ${\hat p}_{\rm interA} $ and $ {\hat p}_{\rm intraA} $. We can see from Table.~\ref{tab:AuthenTab}, the necessary bit length of $ n $ decreases as $ N $ is increasing. For example, $ n $ is reduced by more than 63\% by increasing $ N $ from three to nine whilst both FAR and FRR are guaranteed to be less than 0.001. This validates the high efficacy of the presented ASR selection methodology. Using ASRs that are more sensitive to aging expedite the detecting of recycled commodity SoCs as less response bits need to be acquired during an evaluation. In addition, the volume needed to securely store reference ASRs in database is reduced or relaxed.

Besides the above 48 hrs accelerated aging period, we also test the detection capability given two other accelerated aging periods: 18 hrs and 108 hrs---equal to 8.3 and 49.6 days of SoC operation in the field. The evaluated detection capability is detailed in Table~\ref{tab:DetecTime}. We set $ N=9 $ for all evaluations. We can see that longer aging periods are easier to detect with fewer number of ASRs while guaranteeing the same detection capability, e.g., EER threshold.

In practice, given the same $ n $, if the FAR is more critical than FRR---this maybe the case as FAR poses a security concern by mistakenly accepting recycled SoCs, a smaller $ n_{\rm th} $ can be adopted. 
%For example, given $ N=9 $, $ n=974 $, if we set $ n_{th}= $, we can guarantee the FAR is less than while trading off the FRR to be.
\begin{table*} 
 \centering
 \caption{Quantitative evaluation of necessary bit length of the response for successful detection under different ${\hat p}_{\rm interA} $ and $ {\hat p}_{\rm intraA} $ that are related to aging period, where $ N=9 $.}
 	\begin{tabular}{c c|| p{0.5cm} p{0.5cm} p{0.5cm} p{0.8cm} || p{0.5cm} p{0.5cm} p{0.5cm} p{0.8cm} || p{0.5cm} p{0.5cm} p{0.5cm} p{0.8cm}} % The final bracket specifies the number of columns in the table along with left and right borders which are specified using vertical bars (|); each column can be left, right or center-justified using l, r or c. To specify a precise width, use p{width}, e.g. p{5cm}
 		\toprule % Top horizontal line
 		\toprule % Top horizontal line
 		 &\multicolumn{4}{r}{EER $ < 10^{-2} $} & \multicolumn{4}{r}{EER $ < 10^{-3} $}& \multicolumn{4}{r}{EER $ < 10^{-4} $} \\% Amalgamating several columns into one cell is done using the \multicolumn command as seen on this line
 		\cmidrule(l){3-6} \cmidrule(l){7-10} \cmidrule(l){11-14}% Horizontal line spanning less than the full width of the table - you can add (r) or (l) just before the opening curly bracket to shorten the rule on the left or right side
 		Aging period (Days) &$ {\hat p}_{\rm interA}-{\hat p}_{\rm intraA} $ & $ n $ & $ {\rm n}_{\rm EER} $ & FAR$ ^* $ & FRR$ ^* $ &$ n$ & $ {\rm n}_{\rm EER} $ & FAR$ ^* $ & FRR$ ^* $ & $ n $ & $ {\rm n}_{\rm EER} $ & FAR$ ^* $ & FRR$ ^* $  \\ % Column names row
 		\midrule % In-table horizontal line
		8.3 & 3.32 \% & 1870 & 199 & $ -2.00 $ & $ -2.01 $ & 3294 & 350 & $ -3.00 $ & $ -3.01 $ & 4764 & 506 & $ -4.00 $ & $ -4.01 $ \\ % Content row 1 
		22.1 & 6.52 \% & 551 & 68 & $ -2.01 $ & $ -2.00 $ & 974 & 120 & $ -3.01 $ & $ -3.04 $ & 1406 & 173 & $ -4.01 $ & $ -4.03 $ \\
		49.6 & 8.61 \% & 330 & 43 & $ -2.02 $ & $ -2.01 $ & 584 & 76 & $ -3.01 $ & $ -3.04 $ & 840 & 109 & $ -4.01 $ & $ -4.02 $ \\ % Content row 1
 		\bottomrule % Bottom horizontal line
 		%\bottomrule % Bottom horizontal line
 	\end{tabular}
 	\begin{tablenotes}  
 		\item[a] Note: the $ ^* $ symbol indicates $ \log_{10}({\cdot}) $ of the value.
 		
 	\end{tablenotes}  
 	\label{tab:DetecTime} % A label for referencing this table elsewhere, references are used in text as \ref{label}
 \end{table*}
\section{Conclusion}\label{Sec:Conclusion}
In this study, we experimentally validate the use of embedded SRAMs in off-the-shelf microcontrollers to detect the periods that SoCs work in the field. It is validated that both FAR and FRR can be less than $ 10^{-4} $ when the SoCs experiences only nine days aging. The simplified ASR selection method considerably reduces the necessary number of SRAM PUF response bits to achieve the required detection capability by employing responses that exhibit higher sensitivity to aging effects. In addition, adding the ability of aging sensing to the popular SRAM PUF extends its function to secure IC supply chains by not only preventing cloned and overproduced ICs but also from recycled ones. Most importantly, detection of recycled commercial SoCs embedded with SRAM memories requires no modification to the original design, and thus cost-free is achieved.
\section*{Acknowledgment}
This research was supported by the Australian Research Council Discovery Program (DP140103448). We acknowledge support from China Scholarship Council \\(201306070017). We thank the help from Dr Alex Dinovitser for oven setup and Mr Danny Di Giacomo for experiment setup. We also thank useful discussions with Mr. Zimu Guo and Dr. Domenic Forte.
\newline

%{\bf Reference}
%\bibliography{IEEEfull}

\end{document}